\documentstyle[11pt,equation,
epsf]{article}
\input{psfig.sty}
\begin{document}
\newcommand{\tr}{\rm Tr}
\newcommand{\sx}{\sigma}
\newcommand{\sxa}{\sigma_1}
\newcommand{\sxb}{\sigma_2}
\newcommand{\pha}{\phi_1}
\newcommand{\phb}{\phi_2}
\newcommand{\phpa}{\phi^+_1}
\newcommand{\phpb}{\phi^+_2}
\newcommand{\phm}{\phi^-}
\newcommand{\php}{\phi^+}
\newcommand{\psif}{{\psi_f}}
\newcommand{\psa}{\psi_1}
\newcommand{\psb}{\psi_2}
\newcommand{\Phib}{\bar{\Phi}}
\newcommand{\mpl}{m_{Pl}}
\newcommand{\Mpl}{M_{Pl}}
\newcommand{\lx}{\lambda}
\newcommand{\Lx}{\Lambda}
\newcommand{\kx}{\kappa}
\newcommand{\ex}{\epsilon}
\newcommand{\be}{\begin{equation}}
\newcommand{\ee}{\end{equation}}
\newcommand{\een}{\end{subequations}}
\newcommand{\ben}{\begin{subequations}}
\newcommand{\beq}{\begin{eqalignno}}
\newcommand{\eeq}{\end{eqalignno}}
\def \lta {\mathrel{\vcenter
     {\hbox{$<$}\nointerlineskip\hbox{$\sim$}}}}
\def \gta {\mathrel{\vcenter
     {\hbox{$>$}\nointerlineskip\hbox{$\sim$}}}}
\pagestyle{empty}
\noindent
\begin{flushright}
SNS--PH/1998--3 
\\
hep--ph/9802242
\end{flushright} 
\vspace{3cm}
\begin{center}
{ \Large \bf
Two-Stage Inflation in Supergravity
} 
\\ \vspace{1cm}
{\large 
G. Lazarides$^{(a)}$ and N. Tetradis$^{(b)}$ 
} 
\\
\vspace{1cm}
{\it
(a) Physics Division, School of Technology, University of
Thessaloniki,\\ 54006 Thessaloniki, Greece\\
(b) Dipartimento di Fisica,
Scuola Normale Superiore, \\
56100 Pisa, Italy 
} 
\\
\vspace{2cm}
\abstract{
We investigate the viability of a two-stage inflationary
scenario in the context of supergravity, so as to
resolve the problem of initial conditions for hybrid inflation.
We allow for non-renormalizable terms in the superpotential
and consider the most general form of the
K\"ahler potential and the gauge kinetic function. 
We construct a model
with two stages of inflation, the first 
driven by $D$-term and the second by $F$-term energy
density. 
The viability of this scenario 
depends on the non-minimal terms
in the K\"ahler potential, for which we derive the 
necessary constraints. 
\\
\vspace{1cm}
} 
\end{center}
\vspace{4cm}
\noindent
January 1998

\newpage

\setlength{\baselineskip}{20pt}

\pagestyle{plain}
\setcounter{page}{1}

\setcounter{equation}{0}
\renewcommand{\theequation}{{\bf 1.}\arabic{equation}}

\section{Introduction}

An attractive realization of the inflationary scenario is 
obtained when hybrid inflation 
\cite{hybrid}
is embedded in the context of supersymmetry
\cite{cop}--\cite{gia}.
Flat directions with non-zero potential energy density
appear without fine-tuning and
are not lifted by quantum corrections.
On the contrary, 
the quantum corrections generate 
a small slope of the potential
along the flat directions, which results in the 
slow rolling of the inflaton field $S$.
An important aspect of the hybrid inflationary scenario is 
that the part of inflation with observable consequences 
takes place for values of the inflaton field below the 
Planck scale \cite{cop}
where models can be reliably constructed. 
(Throughout the paper we use the ``reduced'' Planck
scale 
$\mpl= \Mpl/\sqrt{8 \pi}$, $\Mpl = 1.22 
\times 10^{19}$ GeV.)

Recently, the question of initial conditions for hybrid 
inflation has been addressed \cite{first1}--\cite{nikos}. 
In ref. \cite{nikos}, it was shown that 
severe fine-tuning of the initial configuration that will 
lead to inflation is necessary. This is a consequence of the 
presence of (one or more) scalar fields orthogonal to the 
inflaton and the need to satisfy the experimental constraints
resulting mainly from the
COBE observation of the cosmic microwave background
anisotropy.
We briefly summarize the arguments below.

For hybrid inflation, 
the slow rolling of the inflaton occurs along a valley of 
the potential with a small slope. 
In the original scenario \cite{hybrid}, 
the slow rolling  ends when the valley turns into a ridge 
and fluctuations of fields orthogonal to 
the inflaton begin to grow. 
In supersymmetric formulations \cite{cop}--\cite{gia},
it may end even earlier, if the slope along 
the valley becomes large. 
Typically the end of slow rolling
corresponds to a value of the inflaton below the 
Planck scale. The 
COBE observation of the cosmic microwave background
anisotropy constrains the properties of the model along 
the inflationary trajectory \cite{star}--\cite{report}.
On general grounds, one expects the inflationary energy 
scale $V^{1/4}$ 
(determined by the vacuum energy density during inflation) 
to be at least two or three orders of magnitude smaller
than the Planck scale \cite{lyth}:
\be
V^{1/4} / \ex^{1/4} \simeq 7 \times 10^{16}~{\rm GeV}.
\label{oneone} \ee
Here $\ex$ is a ``slow-roll'' parameter \cite{report} 
that must be much smaller than 1 during inflation.

The onset of inflation requires a region of
space with a size of a few Hubble lengths
where the fields take almost constant values, so that the
gradient energy density is negligible compared to the 
potential energy density \cite{gold}. 
The earliest time at which one could start talking about 
such regions of space 
is when the Planck era (during which
quantum gravitational fluctuations dominate) ends and 
classical general relativity starts becoming 
applicable. The initial energy density is near
$\mpl^4$. For a theory with couplings not much smaller 
than 1, the initial field values within each region
are expected to be of order $\mpl$.

Inflation could start at the end of the Planck era, provided 
that the fields take appropriate values. 
However, it is very likely that the 
fields will evolve from some initial values that do not give 
rise to inflation to different values that do.
The difference between the initial energy scale $\mpl$ of 
the field evolution and the inflationary scale $V^{1/4}$ 
implies that the fields evolve for a long time before 
settling down along the inflationary trajectory. The Hubble 
parameter $H$ sets the scale for the ``friction'' term in 
the evolution equations, which determines how fast the energy
is dissipated through expansion. 
When the energy density drops much below $\mpl^4$, the
smallness of the ``friction'' term results in a very long  
evolution, during which the fields oscillate 
many times. Some of the trajectories eventually 
settle down in the valley of the potential that produces 
inflation.
However, the sensitivity to the initial conditions is 
high because of the long evolution. A slight variation of
the initial field values separates inflationary trajectories 
from trajectories that lead to the 
minima of the potential, where inflation does not occur.

This has severe implications for the initial configuration 
that can lead to the
onset of inflation. In ref. \cite{nikos}, it was shown 
that, for the prototype model of
hybrid inflation \cite{hybrid} with a scale consistent 
with the COBE observations, 
the most favourable area of inflationary initial conditions 
is a thin
strip of width $\sim 10^{-5}\mpl$ around the $S$ axis. 
Throughout a region of space with a size of the order of the  
Hubble length (which is 
$\sim \mpl^{-1}$ initially), 
the initial values of the fields orthogonal to the
inflaton must be zero with an accuracy $\sim 10^{-5}\mpl$. 
This should be compared to 
the natural scale of the initial fluctuations of the
fields, which is of order $\mpl$.
If this 
condition of extreme homogeneity is not satisfied the 
fields in different parts of the original
space region will evolve towards very different values.
In one part they may end up in the valley along the $S$ axis,
while in another they may settle at the minima of the 
potential. Before inflation sets in, the 
size of space regions shrinks compared to the
Hubble distance.
As a result, large inhomogeneities are expected at scales 
smaller than
$\sim H^{-1}$ when the evolution of the fields finally
slows down. These will prevent the onset of inflation. 
The fine-tuning of the initial configuration must be 
increased by 
several orders of magnitude (to the $10^{-10}$ level)
if the initial time derivatives
of the fields are non-zero.

The initial condition problem has another serious aspect.
The size $L$ of a homogeneous region evolves 
proportional to the scale factor and, in terms of the 
Hubble length, is given by 
\be
\frac{LH}{L_0H_0} \sim \left( \frac{\rho}{\rho_0} 
\right)^{\frac{1+3w}{6(1+w)}}.
\label{fin} \ee
The parameter $w$ determines the relation between energy 
density and
the mean value of the pressure ($p=w\rho)$. 
For a system of massless oscillating fields, or a 
radiation-dominated Universe,
$w = 1/3$. 
For a system of massive oscillating fields, or a 
matter-dominated Universe,
$w = 0$.
At the onset of inflation, where
$\rho/\mpl^4 \sim 10^{-12}$, 
we must have $LH \gta 1$. 
Otherwise the gradient terms may dominate over the 
vacuum energy density.
For $w=0$, this leads to the requirement 
$L_0 H_0 \gta 100$
at the end of the Planck era, where $\rho_0/\mpl^4 \sim 1$.   
This means that the initial extreme homogeneity must extend 
far beyond the typical size of regions that can be considered
causally connected. 
In ref. \cite{nikos} the evolution of the scale 
factor $R$ relative to the Hubble parameter was studied
numerically, 
starting from an initial value 
$R_0 \sim H^{-1}_0$. At the onset of
inflation, $R$ was found to be 
smaller than $H^{-1}$ typically by a factor 
of order 10--100. This reflects the variation of the effective
value of $w$ between $-1$ and 0 during the initial evolution.

In summary, for hybrid inflation,
one must assume the presence of 
a large number (typically $\sim 10^6$)
of causally disconnected adjacent 
initial regions within which the corresponding fields 
have almost equal values. More specifically, the fields 
orthogonal to the inflationary trajectory must be zero with 
an accuracy at least $\sim 10^{-5}$. Despite the  
difficulty of calculating the probability of such an initial 
configuration, we believe that, under normal circumstances, 
it is highly improbable. In a self-reproducing Universe, 
however, it may become quite probable and the problem of 
initial conditions may not exist. Note that the degree of 
fine-tuning of initial data required in non-inflationary 
cosmology is incomparably worse than the one 
necessary for the onset of inflation.

In ref. \cite{mecostas},
a simple resolution of the issue of fine-tuning
described above was suggested.
A scenario
with two stages of inflation was proposed within the context
of global supersymmetry. 
The first stage has a typical scale
$\sim \mpl$, which implies that the ``friction'' term
proportional to $H$ in the evolution equations of the fields 
is large.
As a result, the system settles down quickly along an almost 
flat direction  of the potential 
and this stage of inflation
occurs ``naturally''. By generating an exponential expansion 
of the initial region of space 
it also provides the homogeneity that 
is necessary for the second stage of inflation. The latter 
has a characteristic
scale much below $\mpl$ and generates the 
density perturbations 
that result in the cosmic microwave background
anisotropy observed by COBE. 
 
It was pointed out, in ref. \cite{mecostas}, that 
the generalization of the two-stage inflationary scenario
in the context of supergravity is difficult. 
The scenario considered there 
involved two hybrid inflations along two $F$-flat directions. 
When global supersymmetry is replaced by supergravity,  
all flat directions are in general lifted and inflation 
becomes impossible.  In the context of canonical supergravity 
(minimal K\"ahler potential) and a linear superpotential of 
the form $W=-\mu^2 S$ (like the one encountered in hybrid 
inflationary models) a cancellation takes place that
prevents the appearance of a mass term for $S$ \cite{cop}. 
Therefore, there is a possibility for $S$ to play the 
role of the inflaton. However, for a superpotential with 
two $F$-flat directions, only a linear 
combination of the possible inflaton fields 
stays ``massless'', with the orthogonal
combination acquiring a large mass term \cite{costas}. 
This implies that only one inflationary stage is likely to 
survive in the context of supergravity. However, this argument 
does not apply to the case where inflation
is driven by a $D$-term
energy density \cite{dterm}. 
The two-stage inflationary scenario 
may then be possible along an appropriate combination of 
$F$-flat and $D$-flat directions. 

The above fine-tuning problem may also be solved 
if a first stage of inflation, at values of the inflaton 
$>m_{Pl}$, 
is incorporated into the scheme. This, however, requires 
tiny coupling 
constants. There is, in principle, no reason why this 
possibility cannot 
be realized in the context of hybrid inflation. Inclusion of 
supergravity, however, makes the discussion of this option 
technically 
difficult. The reason is that, although D-term inflation 
does not contain 
exponentially growing terms at values of the 
inflaton $>m_{Pl}$, a 
reliable expansion scheme is not available in this domain.

In this paper, we discuss a model that allows for 
two stages of inflation, the first driven by
$D$-term and the second by $F$-term energy density. 
We compute the supergravity corrections to the potential
and discuss under what conditions the two stages are
realized. 
In section 2, we introduce our model. We first present
the potential in the globally 
supersymmetric limit and then compute all the relevant 
supergravity corrections.
In sections 3, 4 and 5, we discuss in detail the first 
stage of inflation, the intermediate stage between the 
two inflations and the second stage of inflation 
respectively. We derive constraints for the parameters 
of the K\"ahler potential that guarantee the viability 
of the scenario.
Our conclusions are given in section 6.

\setcounter{equation}{0}
\renewcommand{\theequation}{{\bf 2.}\arabic{equation}}

\section{The model}

\subsection{Global supersymmetry}

We consider the renormalizable superpotential 
\be
W = \lx X \Phi^+ \Phi^- + \kx S \Phi \Phib - \mu^2 S.
\label{four} \ee
The superfields $\Phi$, $\Phib$ are the Standard Model 
singlet
components of a conjugate pair of chiral superfields which 
tranform under the GUT gauge symmetry group 
$G$. Their expectation values break the group,
reducing its rank. 
The superfields $S$, $X$, $\Phi^+$, $\Phi^-$ 
are singlets under $G$.
The parameters $\lx$, $\kx$, $\mu$ can be taken real and 
positive by absorbing their possible phases in a 
redefinition of the fields. 
The superpotential possesses the following 
three $U(1)$ symmetries too: \\
a) $U(1)_R$: a global $R$-symmetry. \\
b) $U(1)_\xi$ : an (``anomalous'') local symmetry 
with gauge coupling 
$g$. \\
c) $U(1)_X$ : a global symmetry. \\
The charges of the fields are

\begin{table} [h]
\renewcommand{\arraystretch}{1.5}
\hspace*{\fill}
\begin{tabular}{|c||c|c|c|c|}	\hline 

& $U(1)_R$
& $U(1)_\xi$
& $U(1)_X$
& $U(1) \subset G$
\\ \hline \hline
  $X$
& 1
& 0
& 1
& 0
\\ \hline 
  $\Phi^+$
& 0
& 1
& $-\frac{1}{2}$
& 0
\\ \hline 
  $\Phi^-$
& 0
& $-1$
& $-\frac{1}{2}$
& 0
\\ \hline
  $S$
& 1
& 0
& 0
& 0
\\ \hline 
  $\Phi$
& 0
& 0
& 0
& 1
\\ \hline 
  $\Phib$
& 0
& 0
& 0
& $-1$
\\ \hline
\end{tabular}
\hspace*{\fill}
\renewcommand{\arraystretch}{1}
\end{table}

Performing appropriate 
$U(1)_R$,
$U(1)_\xi$,
$U(1)_X$ and 
$G$ transformations, and using the vanishing condition
for the $D$-term with respect to $G$ (i.e.
$\left| \Phi \right| =  \left| \Phib \right|$ ),
the fields can be written in the form 
\beq
X =&~\frac{\chi}{\sqrt{2}},
~~~~~~~~~~~~
\Phi^+ = ~\frac{\phpa +i \phpb}{\sqrt{2}},
~~~~~~~~~~~
\Phi^- = ~\frac{\phm}{\sqrt{2}},
\nonumber \\
S =&~\frac{\sigma}{\sqrt{2}},
~~~~~~~~~~~~~~~~~~
\Phi =\Phib = ~\frac{\pha+i\phb}{2}.
\label{five} \eeq
The real fields $\chi$, $\phpa$, $\phpb$, 
$\phm$, $\sx$, $\pha$, $\phb$ 
have canonically normalized kinetic terms.

In the globally supersymmetric limit, 
the scalar potential is given by  
\beq 
V = 
~&\lx^2 \left| X \right|^2 
\left( \left| \Phi^+ \right|^2 + 
\left| \Phi^- \right|^2 \right)
+\lx^2 \left| \Phi^+ \right|^2 \left| \Phi^- \right|^2
+\frac{g^2}{2} \left(
\left| \Phi^+ \right|^2 -\left| \Phi^- \right|^2 + \xi
\right)^2
\nonumber \\
&+\kx^2 \left| S \right|^2 
\left( \left| \Phi \right|^2 
+\left| \Phib \right|^2 \right)
+ \left| \kx  \Phi \Phib  - \mu^2 \right|^2.
\label{fivea} \eeq
It can  be expressed in terms of the real fields defined in
eq. (\ref{five}) as
\beq 
V = 
~&\frac{\lx^2}{4} \chi^2 \left( \left[ \phpa \right]^2 
+ \left[ \phpb \right]^2 + \left[ \phm \right]^2  
\right)
+\frac{\lx^2}{4}  \left( \left[ \phpa \right]^2 + 
\left[ \phpb \right]^2  \right)  \left[ \phm \right]^2 
\nonumber \\
~&+\frac{g^2}{8} \left(
\left[ \phpa \right]^2 +
\left[ \phpb \right]^2 - \left[ \phm \right]^2
+ 2 \xi \right)^2
\nonumber \\
~&+ \left[ \frac{\kx}{4} \left( \pha^2-\phb^2 \right) -
 \mu^2 \right]^2
+ \frac{\kx^2}{4}\pha^2 \phb^2 
+\frac{\kx^2}{4} \sx^2 \left(\pha^2 + \phb^2 \right).
\label{six} \eeq
The minima of the potential are located at 
$\chi=0$, $\phpa =\phpb =0$, 
$\left[ \phm \right]^2=2\xi$, 
$\sx=0$, $\pha^2= 4\mu^2/\kx$, $\phb=0$. 
We choose the Fayet-Iliopoulos
term $\xi$ to be positive and we consider the case 
$g \xi \gg \mu^2$ in the following. 
A Fayet-Iliopoulos term can be generated at the one-loop 
level
if the symmetry $U(1)_{\xi}$ is anomalous \cite{witten}. 
When our model is embedded in the context of a complete 
theory 
(such as string theory) the anomaly is expected to be
cancelled through the Green-Schwarz mechanism 
and $\xi$ is calculable. 
For example, within 
the weakly-coupled heterotic string theory 
it is given by \cite{atick}
\be
\xi = \frac{\tr Q}{192 \pi^2} g^2  \mpl^2, 
\label{fi2} \ee
where 
$\tr Q$ is the total charge 
under the $U(1)_{\xi}$ symmetry.

For $\phpa=\phpb=\phm=\pha=\phb=0$ the potential of 
eq. (\ref{six})
is independent of 
$\sx$, $\chi$ and has the value $V=g^2 \xi^2/2+\mu^4$, 
with the $\mu^4$
contribution being negligible.
Therefore, this range of field values
can support a first stage of inflation. 
The masses of the 
$\phi^+_{1,2}$ and $\phm$ fields are 
\beq
M^2_{\phpa}=M^2_{\phpb}=~&\frac{\lx^2}{2} 
\chi^2 + g^2 \xi.
\nonumber \\
M^2_{\phm}=~&\frac{\lx^2}{2} \chi^2 - g^2 \xi.
\label{seven} \eeq
An instability appears for 
\be
\chi^2 
< \chi^2_{ins} = \frac{2g^2}{\lx^2} \xi,
\label{eight} \ee
which can trigger the growth of the $\phm$ field.

For $\pha=\phb=0$ 
the potential is independent of 
$\sx$. 
The masses of the 
$\phi_{1,2}$ fields are given by 
\beq
M^2_{\pha}=~&\frac{\kx^2}{2} \sx^2-\kx\mu^2, 
\nonumber \\
M^2_{\phb}=~&\frac{\kx^2}{2} \sx^2 +\kx\mu^2.
\label{nine} \eeq
An instability appears for 
\be
\sx^2 
< \sx^2_{ins} = \frac{2}{\kx} \mu^2,
\label{ten} \ee
which can trigger the growth of the $\pha$ field.
For $\chi=\phpa=\phpb=0$, $\left[ \phm \right]^2=2\xi$
the potential is $V=\mu^4$.
Therefore, 
this range of field values can support a second stage 
of inflation.

The flatness of the potential
is lifted by radiative corrections. 
For $\chi \gg \chi_{ins}$, 
the one-loop $\chi$-dependent radiative correction at 
$\phpa=\phpb=\phm=0$ is
\be
\Delta V_r(\chi) = 
\frac{g^4}{16 \pi^2} \xi^2
\left[ \ln\left(\frac{\lx^2 \chi^2}{2\Lambda_1^2} \right) 
\right].
\label{eleven} \ee
For $\sx \gg \sx_{ins}$, 
the one-loop $\sx$-dependent radiative correction at 
$\pha=\phb=0$ is
\footnote{
The radiative correction depends on 
the dimensionality $N$ of the representation of
the GUT group $G$ under which the chiral superfields
$\Phi$, $\Phib$ transform. More specifically, the 
right-hand side of
the one-loop contribution of eq. (\ref{twelve})
must be multiplied by $N$. As we are not specifying the 
group $G$ 
in this work, we have set $N=1$. 
}
\be
\Delta V_r(\sx) = 
\frac{\kx^2}{16 \pi^2} \mu^4
\left[ \ln\left(\frac{\kx^2 \sx^2}{2\Lambda_2^2} \right) 
\right].
\label{twelve} \ee
The precise values of the 
normalization scales $\Lambda_{1,2}$ are 
not important for our discussion. 
Also, away from the flat directions $(\phi^+_{1,2}, \phm,
\phi_{1,2} \not= 0)$ the radiative
corrections are not significant and we neglect them. 

\subsection{Supergravity}

The scalar potential in supergravity has the 
form \cite{sugra}
\be
V= \exp\left( \frac{K}{\mpl^2} \right)
\left[
\left( K^{-1} \right)_i^{~j} F^i F_j 
-3 \frac{\left| W\right|^2}{\mpl^2}
\right]
+ \frac{g^2}{2} {\rm Re}f^{-1}_{AB} D^A D^B,
\label{one} \ee
where 
\beq
F^i=~&W^i+K^i \frac{W}{\mpl^2},
\label{two} \\
D^A=~&K^i \left(T^A \right)^{~j}_i \Phi_j + \xi^A~.
\label{three} \eeq
Here $K$, $W$ and $f$ are respectively the K\"ahler 
potential,
superpotential and gauge kinetic function. 
Upper (lower) indices (i,j) denote differentiation 
with respect to $\Phi_i$ ($\Phi^{j*}$) and $T^A$ 
are the generators 
of the gauge group in the appropriate representation. The 
$\xi^A$ are Fayet-Iliopoulos $D$-terms, which can only 
exist for $U(1)$ gauge groups. 

Allowing for
non-renormalizable terms, 
the most general form of the superpotential permitted 
by the
symmetries discussed in the beginning of the previous 
subsection is 
\be
W=S 
\sum_{n=0}^{\infty} A_n \left( \Phi \Phib \right)^n
+ X \Phi^+ \Phi^-
\sum_{n=0}^{\infty} B_n \left( \Phi \Phib \right)^n,
\label{thirteen} \ee
with $A_0=-\mu^2$, $A_1=\kx$, $B_0=\lx$. 
We are interested in the effect of the supergravity 
corrections on the evolution of the fields during the two
stages of inflation and the intermediate stage.  
These corrections can generate large mass
terms for the inflatons and destroy the slow-roll solutions.
For this reason, we concentrate on the field space near the
flat directions of the potential.
Therefore, the corrections that are relevant for our 
discussion 
have $\Phi=\Phib=\Phi^+=0$. 

The terms in the K\"ahler potential that can contribute
to  $K^i$, $\left( K^{-1} \right)^{~j}_i$
in eq. (\ref{one}) for $\Phi=\Phib=\Phi^+=0$ can be
parametrized as
\beq
K=~&  \sum_{n_1,n_2,n_3=0}^{\infty}
\frac{a_{n_1 n_2 n_3}}
{\mpl^{2 \left( \Sigma n-1 \right)}}
\left| S \right|^{2 n_1} \left| X \right|^{2 n_2}
\left| \Phi^- \right|^{2 n_3}
\nonumber \\
&+ \left(
\frac{1}{\mpl^2} \Phi^+ \Phi^- S^* X
\sum_{n_1,n_2,n_3=0}^{\infty}
\frac{b_{n_1 n_2 n_3}}
{\mpl^{2 \left( \Sigma n \right)}}
\left| S \right|^{2 n_1} \left| X \right|^{2 n_2}
\left| \Phi^- \right|^{2 n_3} + h.c. \right)
\nonumber \\ 
&+ \left| \Phi^+ \right|^2
\sum_{n_1,n_2,n_3=0}^{\infty}
\frac{c_{n_1 n_2 n_3}}
{\mpl^{2 \left( \Sigma n \right)}}
\left| S \right|^{2 n_1} \left| X \right|^{2 n_2}
\left| \Phi^- \right|^{2 n_3}
\nonumber \\
&+ \left| \Phi \right|^2
\sum_{n_1,n_2,n_3=0}^{\infty}
\frac{d_{n_1 n_2 n_3}}
{\mpl^{2 \left( \Sigma n \right)}}
\left| S \right|^{2 n_1} \left| X \right|^{2 n_2}
\left| \Phi^- \right|^{2 n_3}
\nonumber \\
&+ \left| \Phib \right|^2
\sum_{n_1,n_2,n_3=0}^{\infty}
\frac{e_{n_1 n_2 n_3}}
{\mpl^{2 \left( \Sigma n \right)}}
\left| S \right|^{2 n_1} \left| X \right|^{2 n_2}
\left| \Phi^- \right|^{2 n_3},
\label{fourteen} \eeq
with $a_{000}=0$, $a_{100}=a_{010}=a_{001}=1$,
$c_{000}=d_{000}=e_{000}=1$. The coefficients 
$a$, $c$, $d$ and $e$ are real, while the
coefficients $b$ may be complex.

The form of the gauge kinetic function $f$ is 
constrained by its holomorphicity and the symmetries
of the model. The $R$-symmetry and the holomorphicity 
prevent
the appearance of $S$ and $X$ in $f$. The $U(1)_\xi$ 
symmetry 
permits the presence of the combination $\Phi^+ \Phi^-$. 
However
this combination has $X$-charge $-1$, which cannot be 
compensated, as
$X$ cannot appear in $f$. 
The only combination that is allowed
is $\Phi \Phib$. As we are interested 
in the field space with $\Phi=\Phib=\Phi^+=0$ and the 
derivatives 
of $f$ are not relevant for the potential, we conclude that 
we can take $f^{-1}=1$ for our study. 

The potential is given by eq. (\ref{one}).
For $\Phi=\Phib=\Phi^+=0$ it can be parametrized as 
\beq
V=~&\mu^4 
\sum_{n_1,n_2,n_3=0}^{\infty}
\frac{p_{n_1 n_2 n_3}}{\mpl^{2 \left( \Sigma n \right)}}
\left| S \right|^{2 n_1} \left| X \right|^{2 n_2}
\left| \Phi^- \right|^{2 n_3} 
\nonumber \\
&+ \lx^2 \left| X \right|^{2}
\left| \Phi^- \right|^{2}
\sum_{n_1,n_2,n_3=0}^{\infty}
\frac{q_{n_1 n_2 n_3}}{\mpl^{2 \left( \Sigma n \right)}}
\left| S \right|^{2 n_1} \left| X \right|^{2 n_2}
\left| \Phi^- \right|^{2 n_3} 
\nonumber \\
&+ \lx \left| X \right|^{2}  
\left| \Phi^- \right|^{2} \frac{\mu^2}{\mpl^2}
\sum_{n_1,n_2,n_3=0}^{\infty}
\frac{r_{n_1 n_2 n_3}}{\mpl^{2 \left( \Sigma n \right)}}
\left| S \right|^{2 n_1} \left| X \right|^{2 n_2}
\left| \Phi^- \right|^{2 n_3} 
\nonumber \\
&+ \left| X \right|^{2}  
\left| \Phi^- \right|^{2} 
\left( \frac{\mu^2}{\mpl^2} \right)^2
\sum_{n_1=2,n_2,n_3=0}^{\infty}
\frac{s_{n_1 n_2 n_3}}{\mpl^{2 \left( \Sigma n \right)}}
\left| S \right|^{2 n_1} \left| X \right|^{2 n_2}
\left| \Phi^- \right|^{2 n_3} 
\nonumber \\
&+ \frac{g^2}{2} D^2,
\label{fifteen} \eeq
where the generalized $D$-term is given by 
\be
D=~ -
\sum_{n_1,n_2,n_3=0}^{\infty}
n_3 \frac{a_{n_1 n_2 n_3}}
{\mpl^{2 \left( \Sigma n-1 \right)}}
\left| S \right|^{2 n_1} \left| X \right|^{2 n_2}
\left| \Phi^- \right|^{2n_3} + \xi.
\label{sixteen} \ee
Notice that the non-renormalizable terms in the 
superpotential
do not appear in the expression for the potential, even 
for a non-minimal K\"ahler potential. Also, the potential 
depends only
on the magnitudes of the fields despite the contribution 
of terms such as the one in the second line of 
eq. (\ref{fourteen}). 
The coefficients $p_{n_1 n_2 n_3}$, 
$q_{n_1 n_2 n_3}$, $r_{n_1 n_2 n_3}$, 
$s_{n_1 n_2 n_3}$
can be expressed in terms of the coefficients
$a_{n_1 n_2 n_3}$, $b_{n_1 n_2 n_3}$, $c_{n_1 n_2 n_3}$ 
appearing in
the expansion of the K\"ahler potential. 
A lengthy calculation for the terms suppressed by up to
two powers of $\mpl^2$ gives 
\beq
p_{000}=~&1,~~~~~ 
p_{100}=-4~a_{200},~~~~~
p_{010}=1 -a_{110},~~~~~
p_{001}=1 -a_{101},~~~~~
\nonumber \\
p_{200}=~&\frac{1}{2}+16~a^2_{200}-9~ a_{300} 
- 7 ~a_{200},
\nonumber \\
p_{020}=~&\frac{1}{2}+a^2_{110}- a_{120} - a_{110} 
+ a_{020},
\nonumber \\
p_{002}=~&\frac{1}{2}+a^2_{101}- a_{102} - a_{101} 
+ a_{002},
\nonumber \\
p_{110}=~&1 + 8~ a_{200} ~a_{110} + a_{110}^2 
- 4~ a_{210} +2~ a_{110} -4~ a_{200},
\nonumber \\
p_{101}=~&1 + 8~ a_{200}~ a_{101} + a_{101}^2 
- 4~ a_{201} +2~ a_{101} -4~ a_{200},
\nonumber \\
p_{011}=~&1 + 2~ a_{110} ~a_{101} + 
\left| b_{000} \right|^2
-a_{111} - a_{101} -a_{110} + a_{011},
\label{seventeen} \\
q_{000}=~&1,~~~~~ 
q_{100}=1 -c_{100},~~~~~
q_{010}=1 -c_{010},~~~~~
q_{001}=1 -c_{001},~~~~~
\nonumber \\
q_{200}=~&\frac{1}{2}+a_{200}+c^2_{100}-c_{200}-c_{100},
\nonumber \\
q_{020}=~&\frac{1}{2}+a_{020}+c^2_{010}-c_{020}-c_{010},
\nonumber \\
q_{002}=~&\frac{1}{2}+a_{002}+c^2_{001}-c_{002}-c_{001},
\nonumber \\
q_{110}=~&1 + a_{110}+2~c_{100}~c_{010}-c_{110}-c_{010}
-c_{100},
\nonumber \\
q_{101}=~&1 + a_{101}+2~c_{100}~c_{001}-c_{101}-c_{001}
-c_{100},
\nonumber \\
q_{011}=~&1 + a_{011}+2~c_{010}~c_{001}-c_{011}-c_{001}
-c_{010}
+ \left| b_{000} \right|^2,
\label{eighteen} \\
r_{000}=~&- \left( b_{000} + c.c. \right),
\nonumber \\
r_{100}=~&4~b_{000}~a_{200}+b_{000}~c_{100}-2~b_{100}
-3~b_{000} + c.c.,
\nonumber \\
r_{010}=~&b_{000}~a_{110}+b_{000}~c_{010}-b_{010}
-b_{000} + c.c.,
\nonumber \\
r_{001}=~&b_{000}~a_{101}+b_{000}~c_{001}-b_{001}
-b_{000} + c.c.~.
\label{nineteen} \eeq 
The minimal K\"ahler potential corresponds to 
$a_{n_1 n_2 n_3}=0$ for $n_1+n_2+n_3 \geq 2$,
$b_{n_1 n_2 n_3}=0$ for all $n_1,n_2,n_3$,
and $c_{n_1 n_2 n_3}=d_{n_1 n_2 n_3}=e_{n_1 n_2 n_3}=0$ 
for $n_1+n_2+n_3 \geq 1$.
The potential is determined by the coefficients given in the
above equations if only the contributions equal to 1 and 
$\frac{1}{2}$ are kept in the right-hand side.

\setcounter{equation}{0}
\renewcommand{\theequation}{{\bf 3.}\arabic{equation}}

\section{The first stage of inflation}

As we discussed in the introduction, 
inflation does not start immediately after the
Universe has emerged from the Planck era. An initial evolution
of the fields takes place during which they approach an almost
flat direction and eventually settle on a slow-roll trajectory.
We assume that a Robertson-Walker metric is a good approximation 
for the regions of space with uniform fields that we are 
considering. 
The evolution of the fields is given by the standard equations
(overdots denote derivatives with respect to cosmic time)
\beq
\ddot{\phi_i} + 3 H \dot{\phi_i} = &~ 
- \frac{\partial V}{\partial\phi_i},
\label{threeone} \\
H^2 = \left( \frac{\dot{R}}{R} \right)^2 
= &~\frac{1}{3 \mpl^2} \left(
\sum_i \frac{1}{2} \dot{\phi_i}^2 
+ V \right). 
\label{threetwo} \eeq

We have integrated numerically the above equations for
the potential of eqs. (\ref{six}), (\ref{fi2}) with 
$g=0.5$, 
$\xi/\mpl^2 = 10^{-2} $,
$\lx=0.3$,
$\mu/\mpl=10^{-3}$,
$\kx=0.04$. 
The initial values of the fields have been chosen as
$\chi/\mpl=1.1$,
$\phi^+_1/\mpl=0.04$,
$\phi^+_2/\mpl=0.03$,
$\phi^-/\mpl=0.04$,
$\sx/\mpl=1.2$,
$\phi_1/\mpl=0.4$,
$\phi_2/\mpl=0.2$.
They are near $\mpl$, with 
$\phpa$, $\phpb$, $\phm$, $\pha$, $\phb$
smaller than $\chi$ and $\sx$, so that the evolution starts
near the flat direction.
We have also taken into account the 
radiative and supergravity corrections that
provide a small slope along the flat direction. 
To this end, we have included the contributions
of eqs. (\ref{eleven}), (\ref{twelve}) and 
the corrections arising from 
eqs. (\ref{fifteen})--(\ref{nineteen})
with 
$\phi^+_{1,2}=\phi^-=\phi_{1,2}=0$.
Only the parameters $p_{n_1 n_2 n_3}$ are
relevant for our discussion. We have used
values corresponding to a minimal K\"ahler potential:
$p_{100}=0$, $p_{010}=1$, $p_{200}=1/2$,
$p_{020}=1/2$, $p_{110}=1$.  
The contributions 
to the potential from the non-minimal terms
affect mainly the intermediate stage and 
the second stage of inflation
and are discussed in sections 4 an 5. We point out that 
the non-minimal terms 
also modify the kinetic terms in the Lagrangian and,
therefore, the left-hand side of eq. (\ref{threeone}). 
We have studied numerically the resulting corrections to the
solutions presented below
and found that they are small. For this reason, we consider
only the corrections to the potential in the following.

In figs. 1 and 2, we present the evolution 
of the various fields and the Hubble parameter in units of 
$\mpl$. 
The evolution starts at some initial time 
$t \sim H_0^{-1}$, with
$H_0/\mpl \simeq 9.0 \times 10^{-3}$.
(For illustrative purposes we have indicated a time
$t_0/\mpl = 100$ in figs. 1 and 2).
We have taken zero
initial time derivatives for the fields. 
The initial energy density scale is approximately 
one order of
magnitude below $\mpl$ ($\rho_0/\mpl^4 
\simeq 2.5 \times 10^{-4}$).
It is difficult to follow the evolution of the 
system at larger energy
scales, as this typically takes us outside the 
region of validity of 
expansions such as the ones in 
eqs. (\ref{fifteen}), (\ref{sixteen}). 
The initial field values 
$\chi/\mpl=1.1$, 
$\sx/\mpl=1.2$ are already at the border of this region. 
(Notice, however, that the expansion parameters are 
$\left| X \right|^2/\mpl^2 = \chi^2/2\mpl^2$, 
$\left| S \right|^2/\mpl^2 = \sx^2/2\mpl^2$ and they are
always multiplied by powers of $\mu^2/\mpl^2$, 
$\left| \Phi^+ \right|^2/\mpl^2$, 
$\left| \Phi^- \right|^2/\mpl^2$, 
$\left| \Phi \right|^2/\mpl^2$, 
$\left| \Phib \right|^2/\mpl^2$, which we have chosen 
much smaller than 1.)

In fig. 1, 
we observe that 
$\phpa$, $\phpb$ and $\phm$ oscillate around
zero with their amplitude decaying rapidly. 
The field 
$\chi$ quickly settles on the almost flat direction and
approaches a slow-roll solution. 
This behaviour is induced by the large value of 
the ``friction'' term $3 H \dot{\phi}_i$
in the evolution equations of the fields. 
At a  time $t \simeq 800~\mpl^{-1}\sim H_1^{-1}$
the energy density is dominated by the contribution
$g^2 \xi^2 /2$ in the potential and inflation sets in. 
The Hubble parameter during the first stage of inflation
is $H_1/\mpl \simeq 2.0 \times 10^{-3}$.
In fig. 2, we observe that 
$\sx$, $\pha$ and $\phb$ also approach the almost 
flat direction rapidly. 
The slope inducing the slow rolling of $\sx$ is much 
smaller than the one for $\chi$. As a result the 
total 
evolution of $\sx$ during the first stage of inflation is
negligible.  

As we discussed in the introduction, the
onset of inflation requires a region of
space with a size of a few Hubble lengths
where the fields take almost constant values, so that the
gradient energy density is negligible compared to the 
potential energy 
density. It is reasonable to assume that this region 
first emerges at the end of the Planck era, at a 
time $t_{Pl} \sim \mpl^{-1}$. 
During the evolution until inflation
sets in, the size of this region shrinks with respect to the
Hubble length. For the parameters of our model, the 
numerical integration of the evolution equations
indicates that the ratio $R/H^{-1}$ is
reduced by a factor 
4--6 between $t_{Pl}$ and the beginning of inflation.
This implies that a region homogeneous over a Hubble length at
the onset of inflation evolves from a region homogeneous 
over 4--6 Planck lengths at $t_{Pl}$. 
We assume that it is not ``unnatural'' for such a region 
to appear at the end of the Planck era. 
The necessary range of homogeneity at $t_{Pl}$ is reduced 
if the first stage of inflation takes place closer to the 
Planck scale. In our model, this requires larger values of 
$\xi$, which are difficult to reconcile with the 
expansions in eqs. (\ref{fifteen}), (\ref{sixteen}) when
$\phm$ takes the value $\phm=\sqrt{2\xi}$ after the
end of the first stage of inflation. 
In other models, however, this may be possible.
The initial homogeneity 
is expected to be preserved by the short evolution to the 
almost flat direction where inflation starts.
It should be noted that
in the one-stage scenario the size of the homogeneous region 
at $t_{Pl}$ must be assumed
larger by an order of magnitude than in the 
two-stage scenario
(it must extend across 10--100 Planck 
lengths \cite{nikos}).

For our choice of parameters, 
the dominant contribution to the slope along the almost
flat direction for $\chi$ comes from the 
radiative contribution of eq. (\ref{eleven}).
A slow-roll solution exists for
$\chi \gta {\rm{max}}(\chi_{ins}, \chi_r)$, 
where $\chi_r =g\mpl/2 \pi.$
The total number of e-foldings during this stage 
is 
\be
N_1 = \frac{2\pi^2}{g^2}  \frac{\chi^2_0-\chi^2_f}
{\mpl^2},
\label{fourteena} \ee
where $\chi_0$ is the value of the
$\chi$ field when inflation starts and $\chi_f$ the value 
when it ends.
For $\chi_0 \simeq 0.52~\mpl$ and $\chi_f 
= \chi_{ins} \simeq 0.24~\mpl$
(with $\chi_{ins}$ given by eq. (\ref{eight})), we obtain
$N_1 \simeq 17$, which is confirmed by the numerical
solution. We point out that the expression (\ref{eleven})
that we have employed for the determination of $N_1$
is not valid towards the end of the first stage
of inflation. However, the corrections are small because
$N_1$ is dominated by the expansion of the Universe for 
large values of $\chi$. 
As a result of the expansion during the first
stage of inflation, the initial homogeneous region extends
several orders of magnitude beyond the Hubble length when
the intermediate stage starts.  

During the first stage of inflation, almost massless fields 
like $\sx$ have a spectrum of quantum-mechanical fluctuations
characterized by 
\be
\left( \Delta \sx \right)^2_k 
= \left( \frac{H_1}{2 \pi} \right)^2,
\label{threethree} \ee
where $H_1 \simeq 2.0 \times 10^{-3}~ \mpl$ is the Hubble 
parameter. The freezing of fluctuations that cross outside 
the horizon generates classical
perturbations of the fields at superhorizon scales with the 
same spectrum \cite{freeze}.
At the end of the first stage of inflation, 
the energy density in spatial gradient terms 
associated with the perturbations $\sim H_1/2\pi$
of massless fields like $\sx$ is $\sim H^4_1/4\pi^2$ and
provides a negligible contribution to the total energy density.
It is useful to consider also 
the mean square fluctuation of the classical 
field $\sx$ \cite{freeze}
\be
\left( \Delta \sx \right)^2 
= N_1 \left(\frac{H_1}{2 \pi} \right)^2.
\label{threefour} \ee
We obtain $\left( \Delta \sx \right) \simeq
1.3 \times 10^{-3}~\mpl $, which should be 
compared to the mean field value $\sx \simeq 0.49~\mpl$.
We conclude that the $\sx$ field is well approximated by its 
classical value at the end of the first stage of inflation.  
The fluctuations of the massive fields $\phi_i,~i=1,2$ 
generated by the first stage of inflation are given by 
\be
\left( \Delta \phi_i \right)^2_k 
= \left( c~\frac{H_1^2}{M_{\phi_i}} 
\right)^2, 
\label{threefive} \ee
where $M_{\phi_i}$ are their masses, given by
eqs. (\ref{nine}),  and 
$c={\cal O} \left( 10^{-1} \right)$.
In the following sections we shall follow the evolution
of these fluctuations during the intermediate stage 
between the two stages of inflation.

\setcounter{equation}{0}
\renewcommand{\theequation}{{\bf 4.}\arabic{equation}}

\section{The intermediate stage}

When the $\chi$ field rolls beyond its instability point, 
large domains start appearing in which the value of the 
$\phm$ field grows exponentially with time. 
For statistical systems, for which the expansion of the 
Universe is not relevant, 
this process is characterized as spinodal decomposition.
The expansion of the Universe complicates the above picture,
but the details are not important for our discussion. 
We assume that this initial stage of instability is fast and 
soon the fields 
take values away from the $\chi$ axis and 
in the vicinity of  
the minimum at $\chi=0$, $\phm \simeq \sqrt{2\xi}$, where
the curvature of the potential is positive.
Our assumption is reasonable because the $\sx$ field 
rolls to the origin within a time $\sim H^{-1} \simeq 
\sqrt{6} \left( g \xi /\mpl \right)^{-1}$ or slightly 
larger. On the other hand, the 
typical time scale for the growth of the 
$\phm$ field is given by the absolute value of the 
curvature at the origin and is 
$\sim \left( g \sqrt{\xi}\right)^{-1}$. 
As a result, we expect that 
$\phm$ grows to a value near the minimum within 
a fraction of a Hubble time. 
Notice that for $\sx \not= 0$ the
minimum is not located exactly at
$\chi=0$, $\phm = \sqrt{2\xi}$ because of the 
supergravity
corrections to the $D$-term of eq. (\ref{sixteen}).

After a short complicated evolution,
the massive fields $\chi$, $\phm$ are expected to settle 
into a regular oscillatory pattern around the minimum, 
with the Universe 
characterized by an equation of
state $p=w \rho$ with $w\simeq0$. 
This can be verified by calculating the quantity 
$-\dot{H}/H^2=3(1+w)/2$.
Following ref. \cite{juan}, we start with random initial 
values of $\chi$, $\phm$ in the vicinity of the minimum
and examine how fast the dust-dominated era is 
reached. In agreement with ref. \cite{juan}, we find that
$w$ starts from an initial value near $-1$ and 
within less than one e-folding approaches zero.
During the same time 
the total energy density of the Universe is reduced by a 
factor of about 3. 
We do not expect significant changes of 
the value of the $\sx$ field during this time. As 
$w$ is close to $-1$ initially,
we expect that the typical fluctuations of $\sx$
are given by eq. (\ref{threethree}),
and are much smaller than its average value 
$\sx \simeq 0.49~\mpl$ at the end of the first stage of 
inflation. We conclude that, when a regular oscillatory 
pattern of the
$\chi$, $\phm$ fields is established, the energy density 
is $\rho_0\simeq g^2 \xi^2/6$ and the $\sx$ field has a
value $\sx_0\simeq 0.49~\mpl$.

>From this point on, 
the energy density of the oscillating fields 
is dissipated through the expansion of the
Universe or their possible decay into lighter species. 
When the energy density becomes comparable to 
$\mu^4$, the second stage of inflation can begin.
The stability of the $\sx$ field is crucial during this 
intermediate stage. The supergravity corrections of 
eqs. (\ref{fifteen})--(\ref{nineteen}), 
that are proportional to powers of $\sx$, 
can generate a slope along the $\sx$ axis that
may result in the fast rolling of $\sx$ to zero. The 
details depend sensitively on how efficient the transfer 
of energy from the fields $\chi$, $\phm$ to light 
particles is.

\subsection{Absence of decay channels}

We consider first the possibility that the fields 
$\chi$, $\phm$
do not have any decay channels. As a result, they perform 
damped oscillations aroung the minimum 
at $\chi=0$, $\phm \simeq \sqrt{2\xi}$, while their 
energy is dissipated through expansion. 
This generates an effective 
mass for the field $\sx$. 
The largest contributions to this mass come from 
the term proportional to $q_{100}$ in the second line
of eq. (\ref{fifteen}) and from the one proportional to
$a_{101}$ in the expansion of the $D$-term of 
eq. (\ref{sixteen}) around the minimum. For small values 
of $\sx/\mpl$, like the ones we are considering, the 
effective $\sx$ mass has the general form
\be
\left[ m^2_{\sx} \right]_{eff} = fh\frac{\rho}{\mpl^2}, 
\label{effs} \ee
where $f$ is a constant of order 1 and $h$ is equal to 
$q_{100}$ or $a_{101}$ (or a linear combination of them) 
depending on which contribution 
dominates the energy density $\rho$ (or whether they are 
comparable). 
The equation of motion for the $\sx$ field can then be written 
as
\be
\ddot{\sx} + 3 H \dot{\sx} =
- fh\frac{\rho}{\mpl^2}~\sx.
\label{seventeena} \ee
The Hubble parameter is $H=\sqrt{\rho/3\mpl^2}$, while 
the energy density satisfies the equation 
\be
\frac{d\rho}{dt} 
= -\sqrt{3} (1 + w)\frac{\rho^{3/2}}{\mpl}.
\label{eighteena} \ee
The parameter $w$ determines the relation between energy 
density and
the mean value of the pressure over one oscillation 
($p=w\rho)$. 
For a system of massive oscillating fields, such as the one
we are considering, or a matter-dominated Universe, $w = 0$.
For a system of massless oscillating fields with a quartic 
potential or a radiation-dominated Universe,
$w = 1/3$. 
The evolution of $\sx$ as a function of the
energy density is determined by the Euler equation
\be
\rho^2~ \frac{d^2 \sx}{d\rho^2} 
+ \left( \frac{3}{2} - \frac{1}{1+w} \right) 
\rho~\frac{d \sx}{d\rho}
+ \frac{fh}{3(1+w)^2} ~\sx =0.
\label{laone} \ee

We assume that the $\sx$ field is initially at rest.
For  $fh > 3(1-w)^2/16$, 
the solution of the above equation is 
\be
\sx = \sx_0 
\left( \frac{\rho}{\rho_0}\right)^r 
\biggl\{
\cos \left[ s \ln\left( \frac{\rho}{\rho_0}\right) 
\right]
- \frac{r}{s}
\sin \left[ s \ln\left( \frac{\rho}{\rho_0}\right) 
\right]
\biggr\},
\label{latwo} \ee
with 
\beq
r =~& \frac{1-w}{4(1+w)},
\label{lathree} \\
s=~&\sqrt{
\frac{fh}{3(1+w)^2} 
- \left[ \frac{1-w}{4(1+w)} \right]^2 
}.
\label{lafour} \eeq
For  $fh = 3(1-w)^2/16$, 
the solution is
\be
\sx =\sx_0 \left[ 
1- r \ln\left( \frac{\rho}{\rho_0}\right)\right]
\left( \frac{\rho}{\rho_0}\right)^r,
\label{lafive} \ee
with
$r$ given by eq. (\ref{lathree}).
Finally, 
for $fh< 3(1-w)^2/16$, 
the solution is
\be
\sx= \sx_0 \left[
- \frac{r_2}{r_1-r_2}
\left( \frac{\rho}{\rho_0} \right)^{r_1}
+ \frac{r_1}{r_1-r_2}
\left( \frac{\rho}{\rho_0} \right)^{r_2}
\right],
\label{lasix} \ee
with
\be
r_{1,2} = \frac{1-w}{4(1+w)}\pm
\sqrt{
\left[ \frac{1-w}{4(1+w)} \right]^2 
-\frac{fh}{3(1+w)^2}}.
\label{laseven} \ee

During the intermediate stage between the two stages of 
inflation, the energy density drops by a factor 
$\rho/\rho_0 \simeq 6 \mu^4/g^2 \xi^2 \sim 10^{-7}$ 
for the parameters of our model.
For  $fh \gta 3(1-w)^2/16$, the $\sx$ field 
rolls close enough to the origin 
for the second stage of inflation not to take place.
For $ f  \left| h \right| \ll 3(1-w)^2/16$, 
the final value of $\sx$ is approximately given by the 
expression
\be
\sx = \sx_0 \left( \frac{\rho}{\rho_0} 
\right)^{\frac{2fh}{3\left( 1-w^2 \right)}}
\label{appr} \ee
and is close to the original value $\sx_0$. 
Therefore, this parameter range could support a second 
stage of inflation. The value $q_{100}=1$, resulting from
the minimal K\"ahler potential, is too large to
satisfy the above requirement. For $\sx$ not to roll
to the origin and at least 60 e-foldings to be produced 
during the second stage of inflation, 
one must 
assume that the non-minimal corrections fall within the range
$\left| a_{101} \right|, 
\left| q_{100} \right|
= \left| 1-c_{100} \right| \lta 0.1$.
It is not easy to motivate a choice of the non-minimal term  
$c_{100}$ such that the biggest part of the minimal 
contribution to $q_{100}$ is cancelled. 
Another possibility is that the non-minimal corrections take 
values of order 1 and
$a_{101}, q_{100} < 0$. This leads to the rapid growth of
the $\sx$ field during the intermediate stage. One must 
then  
rely on higher non-minimal terms in order to stabilize 
the value of $\sx$
and keep it sufficiently below $\mpl$ for the supergravity 
corrections to be under control.

\subsection{Production of light particles}

The most efficient way of keeping
$\sx$ constant during the intermediate stage is 
through the trasformation
of the energy of the $\chi$, $\phm$ fields into 
lighter particles. 
This is easily achieved if we
introduce a light superfield $\Psi$, singlet under
the MSSM and
with charges (0,0,$-1/2$,0) under the symmetries of 
subsection 2.1. In the globally supersymmetric limit, the
only allowed new term in the renormalizable superpotential 
is $\nu X \Psi^2$. We assume that the coupling
constant $\nu$ is of order 1. The new term is not 
expected to
modify the first stage of inflation, because
the scalar component of $\Psi$ has a zero expectation value.

In our model we have chosen couplings such that the 
masses of the $\php_{1,2}$, $\phm$, $\chi$ particles 
near the
minimum at $\chi=0$, $\phm \simeq \sqrt{2\xi}$ satisfy
$M_{\phm}=\sqrt{2}g \sqrt{\xi} >
2 M_{\chi}=2 M_{\php_{1}}=2 M_{\php_{2}}
=2 \lx \sqrt{\xi}$. As a result, 
several channels exist for the transformation 
of the energy of the
$\chi$, $\phm$ fields into lighter particles.
Although
this process may be accelerated by a period of 
preheating \cite{juan}, we shall not consider this 
possibility 
here, as the standard decay rate is very efficient for our 
purposes.
The various decay channels of the fields are mainly 
determined 
by the Lagrangian of the globally supersymmetric theory. The 
supergravity corrections are less efficient, as they are 
suppressed by powers of $\mpl$. 
The $\phm$ field decays mainly into $\php_{1,2}$ and 
$\chi$ scalar particles and their fermionic partners. 
Subsequently, the $\php_{1,2}$ particles decay into 
light scalar $\Psi$ particles, the 
$\chi$ particles 
decay into fermionic $\Psi$, while the fermionic 
$X$, $\Phi^{+}$ 
decay into fermionic and scalar $\Psi$ particles. 
The $\chi$ field 
oscillates around a zero expectation value and 
decays only into fermionic $\Psi$ directly.
The various decay rates 
have the 
general form
\be
\Gamma = \frac{G}{16 \pi} \sqrt{\xi},
\label{decay} \ee
where $G$ is a function of the couplings
$\lx$, $g$ and $\nu$.
For example, for the decay rate of the $\phm$ field into 
$\php_1$ particles 
\be
G = \frac{\left(g^2-\lx^2\right)^2}{\sqrt{2}g}
\sqrt{1-\frac{2\lx^2}{g^2}}.
\label{decaya} \ee

The total energy density converted into light 
particles at the end of reheating 
is approximately given by \cite{ct}
\be
\rho_r \sim \mpl^2 \Gamma^2_{eff}.
\label{rho} \ee 
Here $\Gamma_{eff}$ is approximately equal to the total 
decay rate of the $\chi$, $\phi$ fields if they decay at 
comparable time scales, or equal to the 
smallest decay rate if the two fields decay at different 
time scales without efficient energy exchange during their 
oscillations. For our model
we expect $\rho_r$ to be about one order of
magnitude smaller than
the energy density at the beginning of the oscillatory 
stage $\rho_0 \simeq g^2 \xi^2/6$. 
The solution of eq. (\ref{latwo}) 
implies that the value of the $\sx$ field is
reduced by an approximate factor of 2--3 during the 
decay of the oscillating fields.

Within the globally supersymmetric theory, 
the $\sx$ field has no interaction with the gas of
$\Psi$ particles whose energy density is dominant after 
the end of the oscillatory phase. 
However, supergravity 
introduces additional terms in the potential. 
The most important supergravity correction for our 
discussion
arises when the term $\sim \nu^2 \left| \Psi \right|^4$ 
coming from the $F$-term is multiplied
by a correction $\sim \left|S\right|^2/\mpl^2$ 
originating in the K\"ahler potential.
In analogy to
the corrections of eq. (\ref{fifteen}),
the resulting term has the form 
$\nu^2 \left|\Psi \right|^4 
t_{100} \left|S \right|^2/\mpl^2$, 
with $t_{100}$ a new parameter of order 1.
There are also interaction terms between the $\sx$-field 
and the fermionic $\Psi$ particles. However, they involve 
also the heavy $\chi$ particles or their fermionic partners 
which are not expected to be abundant. For this reason 
we conclude that the $\sx$
field interacts mainly with the gas of the 
scalar $\Psi$ particles. 

The scalar 
$\Psi$ particles have a large self-interaction rate and
their gas is expected to reach thermal equilibrium quickly. 
However, the
interactions of the fermionic $\Psi$ involve additional 
heavy $\chi$ particles and their superpartners,
and thermalization is more difficult for them. 
The interaction of the $\sx$ field with the thermal
bath of the $\Psi$ particles is established through the 
term $\nu^2 \left|\Psi \right|^4
t_{100} \left|S \right|^2/\mpl^2 $. 
The largest contribution to the 
effective mass of $\sx$ is obtained from this term if the 
$\Psi$ legs are contracted in pairs. The thermal part of 
the resulting two-loop graph gives
\be
\left[ m^2_{\sx} \right]_{eff} \sim
\nu^2 t_{100} \left( \frac{T^2}{12} \right)^2 
\frac{1}{\mpl^2}.
\label{effst} \ee

The temperature $T$ can be expressed in terms of the energy 
density. If both the scalar and fermionic $\Psi$ particles 
were in thermal equilibrium the total energy density would be
\be
\rho= g_* \frac{\pi^2}{30} T^4, 
\label{rhot1} \ee
with an effective number of degrees of freedom
$g_*=2 + (7/8)2=3.75$.  
If the fermionic $\Psi$ do not thermalize the total 
energy density is
\be
\rho > g_* \frac{\pi^2}{30} T^4, 
\label{rhot2} \ee
with $g_*=2$. 
In any case, the effective mass of the $\sx$ field
is given by eq. (\ref{effs}), where $h=\nu^2 t_{100}$ 
is of order 1 and $f \lta 10^{-2}$. 
The evolution of the $\sx$ field during the 
radiation-dominated phase is 
given by eq. (\ref{appr}) with $w=1/3$ and  
$\rho_0$ equal to the energy density at the end of the
oscillatory phase $\rho_r$.
The change of $\sx$ by the end of the intermediate stage 
when $\rho \sim \mu^4$ is expected
to be less than 10\% of its value at the end of the 
oscillatory phase.

We conclude that, within the scenario that permits the 
decay of the $\chi$, $\phm$ fields into light particles, 
the value of the
$\sx$ field is reduced by an approximate factor of 
2--3 during the whole intermediate stage. 
As the $\sx$ field does not roll to zero
during this stage, there is no need for 
an ``unnatural'' choice of the 
non-minimal terms in the K\"ahler potential as in the 
previous subsection. In section 3, we computed a value
$\sx \simeq 0.49~\mpl$ at the end of the first stage of
inflation. We shall use an initial value
$\sx \simeq 0.2~\mpl$ for the study of the second stage 
of inflation in the following section.

\setcounter{equation}{0}
\renewcommand{\theequation}{{\bf 5.}\arabic{equation}}

\section{The second stage of inflation}

At the end of the intermediate stage, the fields $\sx$, 
$\phi_{1,2}$ are located on the flat direction of the 
potential where  $V=\mu^4$. 
When the energy density falls below $\mu^4$, the second 
stage of inflation can begin. 
Due to the rapid expansion during the first stage of 
inflation, the homogeneous regions extend far beyond the 
typical Hubble length $H^{-1}_2$ of the second stage. 
However, the size of the fluctuations of the 
$\sx$, $\phi_{1,2}$ fields that were
generated during the first stage must be calculated 
with care, as they can prevent the onset of inflation.

The fluctuations of the massless field $\sx$
that cross inside the horizon during the intermediate stage
start propagating as massless 
particles and their energy density drops $\sim R^{-4}$. 
As a result, we do not expect a significant contribution 
to the total energy density from them.

The fluctuations of the massive $\phi_{1,2}$ fields, that 
were generated by the
first stage of inflation, are given by 
eq. (\ref{threefive}).
During the intermediate stage, the amplitude of these
fluctuations drops as $R^{-3/2}$. As a result 
we expect that, when
the energy density becomes comparable to 
$\mu^4$, this amplitude is approximately given by 
\be
\left( \Delta \phi_i \right)_k
\simeq c~\frac{H^2_1}{M_{\phi_i}} 
\left( \frac{H_2}{H_1} \right)^{\frac{1}{1+w}},
\label{threesix} \ee
with $i=1,2$.
Here $H_1 \simeq 2.0 \times 10^{-3}~ \mpl$
and $H_2 \simeq 5.8 \times 10^{-7}~ \mpl$
are the values of the Hubble parameter during the 
first and second
stage of inflation respectively,
$w$ is determined by the equation of state during the 
intermediate stage and
varies between 0 and 1/3, 
 $M_{\phi_i}$ are given by
eqs. (\ref{nine}), and 
$c={\cal O} \left( 10^{-1} \right)$.

It was shown in ref. \cite{nikos} that for
$\mu \lta 10^{-1} \mpl$
the most favourable area of inflationary initial conditions 
is a thin
strip around the $\sigma$ axis. If the fields start in 
this area without
initial time derivatives $\sx$ does not oscillate around 
zero, but quickly settles along the flat direction.
We can obtain a rough estimate of the width of this area 
if we consider the equation of motion of the $\sx$ field 
and replace $\pha^2$ and $\phb^2$ by their
average values $\langle \phi_{1,2}^2 \rangle
\sim \left[ \phi_{1,2} \right]^2_0/2$ during the 
evolution 
($\left [\phi_{1,2} \right]_0$ are the amplitudes of 
$\phi_{1,2}$).
The equation reads
\be
\ddot{\sx} + 3 H \dot{\sx} =  
- \frac{\kx^2}{4} 
\left( \left[ \pha \right]^2_0 
+ \left[ \phb \right]^2_0
\right) \sx.
\label{twothirteen} \ee

Two time scales
characterize the solutions of this equation.
The first one is related to the ``friction'' term 
and is given by  
$t_H^{-1} = 3H/2$. For 
$\left[ \phi_{1,2} \right]_0 \ll 2 \mu/\sqrt{\kx}$ 
and 
$\kx^2 \sx^2 \left( \left[ \pha \right]^2_0 
+ \left[ \phb \right]^2_0
\right)/8 \ll \mu^4$,
we have 
\be
t_H =  \frac{2}{\sqrt{3}}  \frac{\mpl}{\mu^2}.
\label{twofourteen} \ee
The other time scale is obtained if we neglect the 
``friction'' term and consider the oscillations 
of the $\sx$ field.
One-fourth of the period is the typical time for the 
system to roll
to the origin and away from an inflationary solution. 
It is given by 
\be 
t_{osc} = \frac{\pi}{\kx} 
\frac{1}{\sqrt{\left[ \pha \right]^2_0 
+ \left[ \phb \right]^2_0}}.
\label{twofifteen} \ee
Inflation sets in if $t_{osc} \gta t_H$, which gives 
\be
\frac{\sqrt{\left[ \pha \right]^2_0 
+ \left[ \phb \right]^2_0}}{\mpl} 
\lta \frac{\sqrt{3}~\pi}{2~\kx} 
\left( \frac{\mu}{\mpl} \right)^2.
\label{twosixteen} \ee
We have verified numerically that the above relation 
gives the correct 
order of magnitude for the size of the strip around the 
$\sx$ axis that
leads to inflationary solutions. 
Our assumptions for the derivation of the above bound break 
down when 
$\sx \gta \sqrt{32/3\pi^2}  \mpl \simeq \mpl$.

We can approximate the initial amplitudes  
$\left[ \phi_{i} \right]_0$ by
the typical scale-invariant fluctuations at the end of the 
intermediate stage,
given by eq. (\ref{threesix}). 
For the parameters of our model, the bound of 
eq. (\ref{twosixteen}) 
is comfortably satisfied. As a result, the $\sx$ field is 
expected to settle quickly on a slow-roll trajectory.  
We conclude that the necessary conditions for the onset of 
the second stage of inflation are ``naturally'' satisfied 
in our model.

During the second stage of inflation, 
the main contributions to the slope along the 
$\sx$ direction come from the 
term 
$\mu^4~p_{100}~\sx^2/2 \mpl^2$ 
in eq. (\ref{fifteen}) and the radiative correction of
eq. (\ref{twelve}). Higher-order corrections are 
suppressed
by powers of $\sx^2/2\mpl^2 \lta 2 \times 10^{-2}$ and
$\xi/\mpl^2 =  10^{-2}$. The existence of a 
slow-roll solution requires $p_{100}$ to be smaller than 1.
According to eqs. (\ref{seventeen}) 
this translates into a constraint for the parameter 
$a_{200}$
of the non-minimal K\"ahler potential, which must be 
negative and $\left| a_{200} \right| \lta 10^{-1}$.
However, if one demands that the slow-roll 
solution is driven by the radiative contribution of eq.
(\ref{twelve}) one must choose a much smaller value.
The values of the other parameters are not
constrained for sufficiently small 
$\sx/\mpl$. As in our scenario
$\sx/\mpl \lta 0.2$ during the second stage of inflation,
we assume that they are of order 1 in general. However,
for scenarios of inflation with field values closer to
$\mpl$ a significant fine-tuning of an infinite number of 
parameters is required for the flatness condition to be
maintained by the supergravity corrections. 
For example, a small value of $p_{200}$ in 
eqs. (\ref{seventeen}) can be obtained
only if a carefull cancellation is arranged involving
the factor $1/2$ 
arising from the minimal K\"ahler potential
and the non-minimal parameters $a_{200}$, $a_{300}$.

Because of the supergravity corrections to the $D$-term of 
eq. (\ref{sixteen}), its minimum is a function of the 
$\sx$ field. However, no significant contributions 
to the slope along the inflationary trajectory are generated 
by this term. The reason is that such contributions are 
proportional to the value of the $D$-term, which is very 
rapidly adjusted to zero through a change of the expectation 
value of $\phm$ as $\sx$ 
slowly rolls down the $\sx$ axis. We have verified 
numerically 
this very rapid adjustment of the $D$-term to zero. 

The second stage of inflation terminates when $\sx$ rolls 
below either the instability 
point of eq. (\ref{ten}), or the value $\sx_r = 
\kx \mpl/\sqrt{8 \pi^2}$
for which the logarithmic contribution to the
potential destroys the slow-roll solution.
The number of $e$-foldings during the second stage of 
inflation is 
\be
N_2 = \frac{1}{2p_{100}} \ln \left( 
\frac{\frac{\sx_0^2}{\mpl^2}
+\frac{\kx^2}{8\pi^2}\frac{1}{p_{100}}}
{\frac{\sx_{f}^2}{\mpl^2}
+\frac{\kx^2}{8\pi^2}\frac{1}{p_{100}}} \right), 
\label{lathirteen} \ee 
with $\sx_0$ the value of $\sx$ at the beginning of this 
stage and 
$\sx_f$ the largest of $\sx_{ins}$, $\sx_{r}$.
For the parameter range of interest to us, the above 
expression can be approximated by
\be
N_2 \simeq \frac{1}{p_{100}} \ln \left( 
\frac{2 \pi}{\kx}\sqrt{2 p_{100}} \frac{\sx_0}{\mpl} 
\right).
\label{lathirteena} \ee 
The most stringent constraint on $p_{100}$
results from the need to generate at least 60 e-foldings 
during the
second stage of inflation. If this were not the case, the 
predicted spectrum of adiabatic density perburbations would 
depend
on the parameters of the first stage of inflation and would 
be incompatible with the COBE 
measurements of the cosmic microwave background
anisotropy. For $\sx_0$ of order $\mpl$ and taking $\kx$ 
an order of magnitude smaller than $\sx_0/\mpl$, we find 
$p_{100} \lta 0.05$. This results in the constraint 
$\left|a_{200}\right|\lta 10^{-2}$,
consistently with the results of ref. \cite{costas}. 
(Our parameter $a_{200}$ corresponds to 
$-\beta/4$ of ref. \cite{costas}.)

The spectrum of the adiabatic density perturbations 
generated by this stage of inflation is 
\be
\delta_H^2 \simeq \frac{1}{75 \pi^2}
\frac{1}{(p_{100})^2} \frac{\mu^4}{\mpl^2 \sx^2_{60}}, 
\label{lafourteen} \ee
where $\sx_{60}$ is the value of the $\sx$ field
at which our present 
horizon scale crossed outside the inflationary horizon. This 
value corresponds to $N_Q \simeq 60$ e-foldings.
Comparison with the value $\delta_H=1.94 \times 10^{-5}$, 
deduced from the COBE observation of the cosmic microwave 
background anisotropy, leads to the constraint
\be
\frac{\mu}{\mpl} \simeq  7.7 \times 10^{-3} 
\kx^{1/2} (p_{100})^{1/4} e^{30 p_{100}}.
\label{lasixteen} \ee
In our model, 
we have taken $\kx=0.04$, 
$\mu/\mpl=10^{-3}$ and obtained
$\sx_0/\mpl \simeq 0.2$ at the beginning of the second 
stage of inflation. 
For the choice 
$p_{100}=0.03$, $a_{200}=-7.5 \times 10^{-3}$,
we find $N_2 = 68$ e-foldings and
obtain approximate agreement with the COBE observations.

\setcounter{equation}{0}
\renewcommand{\theequation}{{\bf 6.}\arabic{equation}}

\section{Conclusions}

In this paper, we addressed the problem of 
fine-tuning of the initial conditions for hybrid inflation 
in the context of supergravity. 
This problem is generated by 
the difference between the energy scale at which the 
Universe 
emerges from the Planck era (near $\mpl$) 
and the inflationary scale implied by the
COBE observations ($V^{1/4} \sim 10^{-3}\mpl$).
A simple resolution of this issue of fine-tuning
can be obtained in a scenario
with two stages of inflation at two different energy scales
\cite{mecostas}.
The first stage has a typical energy scale
not far from $\mpl$. 
As a result, the Hubble parameter 
stays large until the fields settle down along the direction 
that produces inflation. Due to the large ``friction'' term
proportional to $H$
in the equations of motion, the initial part of the 
evolution towards the inflationary trajectory 
is short and the first stage of inflation 
occurs ``naturally''. 
After the end of this stage a subset of the fields moves 
towards the minimum of the potential around which it 
performs damped oscillations.  
During this intermediate 
stage, the energy density is reduced 
through expansion. A second stage of inflation 
begins when the energy density
falls below the false vacuum energy density associated with
a second-order phase transition involving the
remaining fields.  
The homogeneity far beyond the Hubble length, that was 
produced during the first inflationary stage, 
makes the onset of the second stage ``natural'', despite 
the fact that its Hubble length is much larger than the 
one of the first stage. 
The second stage of inflation generates the density 
perturbations that result in the cosmic microwave 
background anisotropy observed by COBE.

The realization of the above scenario
in the context of supergravity must overcome two main 
obstacles. Firstly, it is difficult to preserve 
flat directions of the potential in supergravity. 
Typically, fields that are massless at the globally
supersymmetric level develop masses of 
the order of the Hubble parameter that prevent the
onset of inflation. 
A notable exception is a $D$-flat direction, which is 
typically preserved in supergravity \cite{gia}. 
Another exception is an $F$-flat direction in models with
a linear superpotential of 
the form $W=-\mu^2 S$. 
For a minimal K\"ahler potential, this flat direction is 
not lifted in the context of supergravity \cite{cop}.
In this paper, we considered a scenario with two
stages of inflation, the first driven by $D$-term energy
density and the second by $F$-term energy density
resulting from a superpotential of the form 
$W=-\mu^2 S$.
One would expect that a 
scenario with two stages of $D$-term inflation could pose 
fewer technical problems. However, 
the Fayet-Iliopoulos term that provides the 
$D$-term energy density can be naturally generated by 
invoking an ``anomalous''
$U(1)$ symmetry in the effective theory resulting from 
string theory \cite{witten,atick}. Such a term is expected 
to have a scale not far from $\mpl$. 

We calculated the 
potential in our model, allowing for all possible 
non-renormalizable terms in the superpotential 
and considering the most general form of the
K\"ahler potential and the gauge kinetic function.
We found that the potential depends only on 
the non-minimal terms of the 
K\"ahler potential.
We also checked that these terms generate  corrections to 
the kinetic part of the Lagrangian that
do not modify our scenario and can be neglected. 
For the supergravity corrections not to destroy the 
flat directions of the globally supersymmetric model and 
inflation to set in, 
the parameters of the K\"ahler potential must be
chosen appropriately. 
If inflation driven by $F$-term energy density takes 
place for field values near $\mpl$, an
infinitely large number of such parameters must be tuned
in order for the flatness conditions to be satisfied.
However, in our scenario we arranged for inflation to take 
place for sufficiently small field values, so that only one 
condition is necessary:
The coefficient $a_{200}$ of the $\left| S \right|^4$ 
term in the K\"ahler potential, where $S$ is
the inflaton of the second stage, must be chosen 
negative and $\left| a_{200} \right| \lta 10^{-1}$. 

The second difficulty faced by two-stage inflation
is to keep the inflaton $S$ of the second stage essentially 
constant at a value with non-zero $F$-term energy 
density during the intermediate stage.
An effective mass for $S$ is generated through
its couplings in the K\"ahler potential
with the set of fields that after the 
first stage of inflation move to a minimum of the potential 
and oscillate around it. We found that, if these fields have 
no decay channels into lighter particles, some couplings
must be carefully adjusted to values that are difficult 
to justify.
However, if the set of oscillating fields can decay into 
light species, these constraints are not necessary. 
The reason is that $S$ has a weak coupling to the 
generated gas of particles, and its effective mass is 
small enough for it to remain on the flat direction that 
leads to the second stage of inflation.  

The most stringent constraint on the parameters of the model
results from the need to generate at least 60 e-foldings 
during the second stage of inflation. If this were not the 
case, the predicted spectrum of adiabatic density 
perburbations would depend
on the parameters of the first stage of inflation and would 
be incompatible with the COBE 
measurements of the cosmic microwave background
anisotropy. 
As we arranged for inflation to take place for 
sufficiently small field values, only one condition is 
necessary. The coefficient 
$a_{200}$ of the non-minimal $\left| S \right|^4$ term
in the K\"ahler potential
that contributes to the slope along the 
inflationary trajectory of the second stage 
must be 
negative with $\left| a_{200} \right| \lta 10^{-2}$. 
This range of values is narrower by an order of magnitude
than the range that satisfies the slow-roll conditions for 
inflation. It can be justified within an effective field 
theory, if we view the non-minimal terms 
as higher-order corrections to
the minimal form of the K\"ahler potential.

We conclude that a two-stage inflationary scenario can
be realized for carefully
chosen models in the context of supergravity, without
the need to fine-tune an infinite number of parameters.
In the model we presented, only one parameter must
be mildly tuned.
Within a two-stage inflationary scenario the problem of
initial conditions of hybrid inflation can be resolved. 

\vspace{0.5cm}
\noindent
{\bf Acknowledgements}: 
We would like to thank G. Dvali for suggesting the 
possibility of a $D$- and $F$-term inflation and 
for many useful discussions. We would also like to
thank Z. Berezhiani and D. Comelli for useful discussions.
This research was supported by the E.C.
under TMR contract No. ERBFMRX--CT96--0090.

\newpage

\pagestyle{empty}

\begin{figure}
\psfig{figure=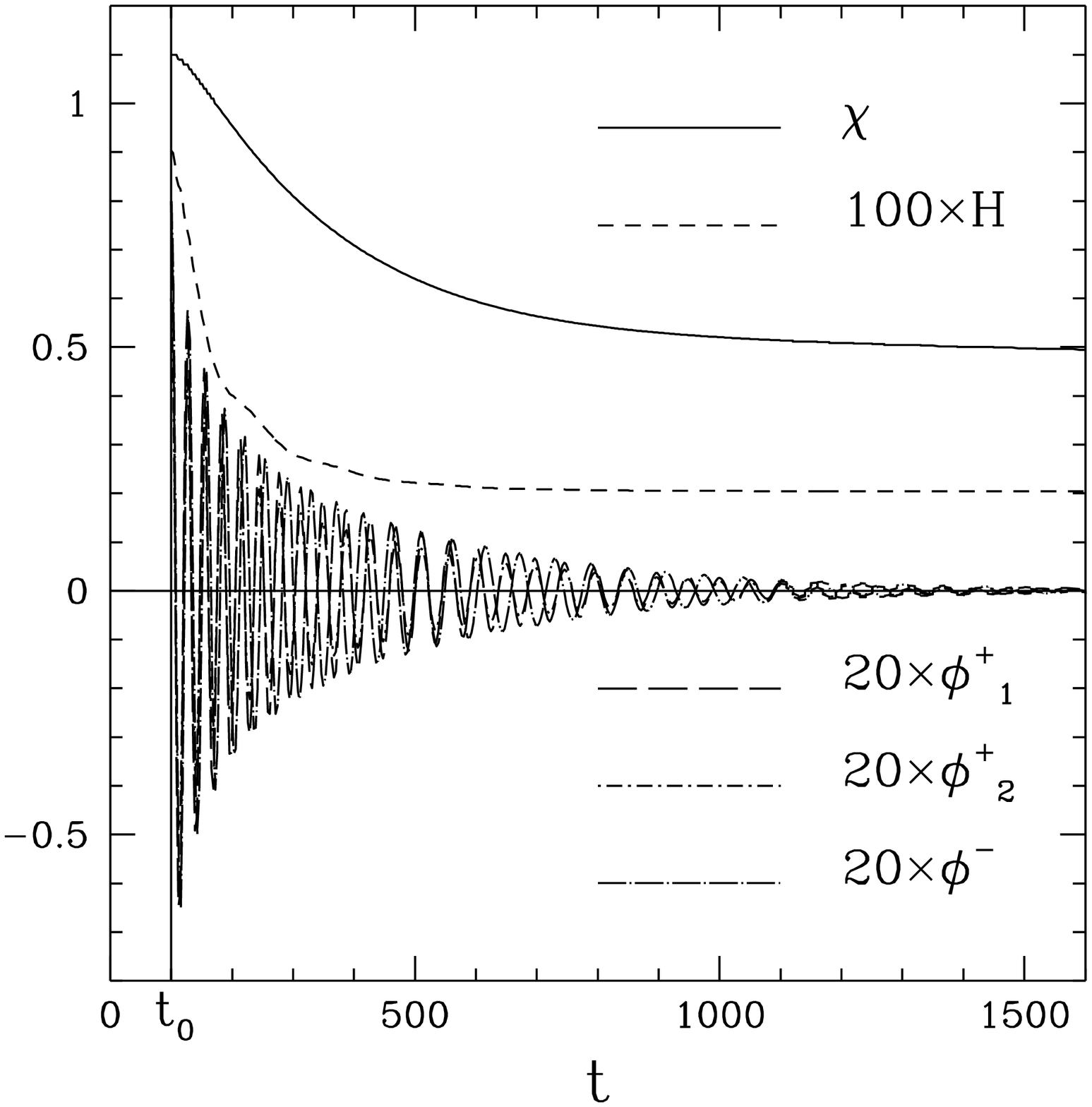,height=16.0cm}
Fig. 1: 
The evolution of 
$\chi$, $\phi^+_1$, $\phi^+_2$, $\phi^-$ 
and the Hubble parameter $H$ for a theory described by
the tree-level potential of eq. (\ref{six}) with 
$g=0.5$, 
$\xi/\mpl^2 = 10^{-2} $,
$\lx=0.3$,
$\mu/\mpl=10^{-3}$,
$\kx=0.04$. The initial values of the fields have been 
chosen as
$\chi/\mpl=1.1$,
$\phi^+_1/\mpl=0.04$,
$\phi^+_2/\mpl=0.03$,
$\phi^-/\mpl=0.04$,
$\sx/\mpl=1.2$,
$\phi_1/\mpl=0.4$,
$\phi_2/\mpl=0.2$.
Dimensionful quantities are given in units of $\mpl$.
\end{figure}

\begin{figure}
\psfig{figure=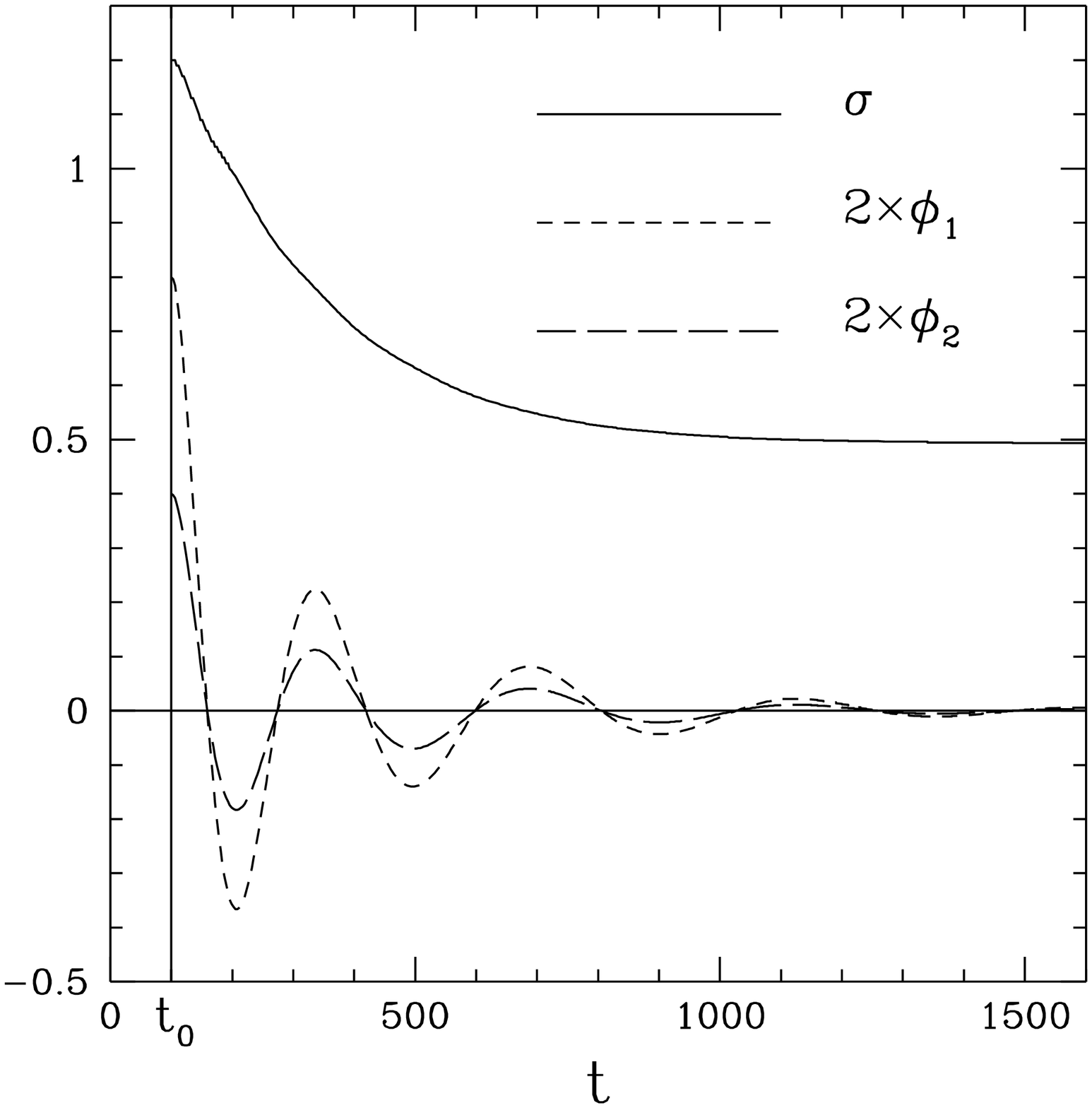,height=16.0cm}
Fig. 2: 
Same as in fig. 1 for 
$\sigma$, $\phi_1$, $\phi_2$.
\end{figure}

\end{document}